\input{psfig.sty}
\documentstyle[11pt,newpasp,twoside]{article}
\def\edcomment#1{\iffalse\marginpar{\raggedright\sl#1\/}\else\relax\fi}
\reversemarginpar

\begin{document}
\title{Spectra and evolution of two X-ray sources in M81}
 \author{V. La Parola, G. Peres}
\affil{DSFA - Sez. di Astronomia, Piazza del Parlamento 1, 90134 Palermo, Italy}
\author{G.Fabbiano, D. W. Kim}
\affil{Harvard Smithsonian Center for Astrophysics, 60 Garden st, Cambridge, MA
02138, USA}

\begin{abstract}
We analyzed the spectral and temporal features of 
X-ray sources in M81 using data from all the relevant
ROSAT-PSPC/HRI observations.  We discuss
the main features of the point-like nucleus and of the second
brightest source in the field (X-9), whose identification is still unclear.
\end{abstract}

\noindent
{\bf M81 nuclear source}\\
M81 has a LINER nucleus, consistent with an X-ray
point-like source (Dyson, priv. comm.). We show its lightcurve over the $\sim$7
years monitored by ROSAT (Fig. 1): each point
is one observation (La Parola et al., in prep.). We also 
looked for spectral variability through hardness ratio (HR) analysis and found
that the source has three spectral states: we analysed the three
spectra derived adding observations with similar HR. 
The best spectral fit for the hard phase (HR $>$ 
0.21) is a two-T Raymond-Smith thermal model(kT$_1 = 0.23\pm0.01$
KeV, kT$_2 = 2.4\pm0.4$ keV); we can model the two
softer phases with a power law + Raymond-Smith with
compatible values($\Gamma\sim2.1$, kT$\sim0.4$ keV). Luminosity is $\sim2\times
10^{40}$ ergs/sec.
\begin{figure}[ht]
\centerline{\psfig{figure=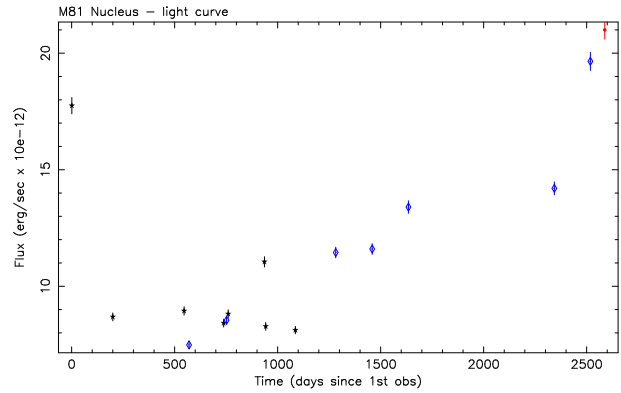,width=9.3cm,angle=270}}
\caption{Light curve of M81 nucleus. Each point is the average flux in one
observation: stars - PSPC, diamonds - HRI, circle - SAX. All the instruments
agree well and strong
variability is present on short (days) and on long (years) time scales.}
\label{nflux}
\end{figure}
\begin{figure}[h]
\centerline{\psfig{figure=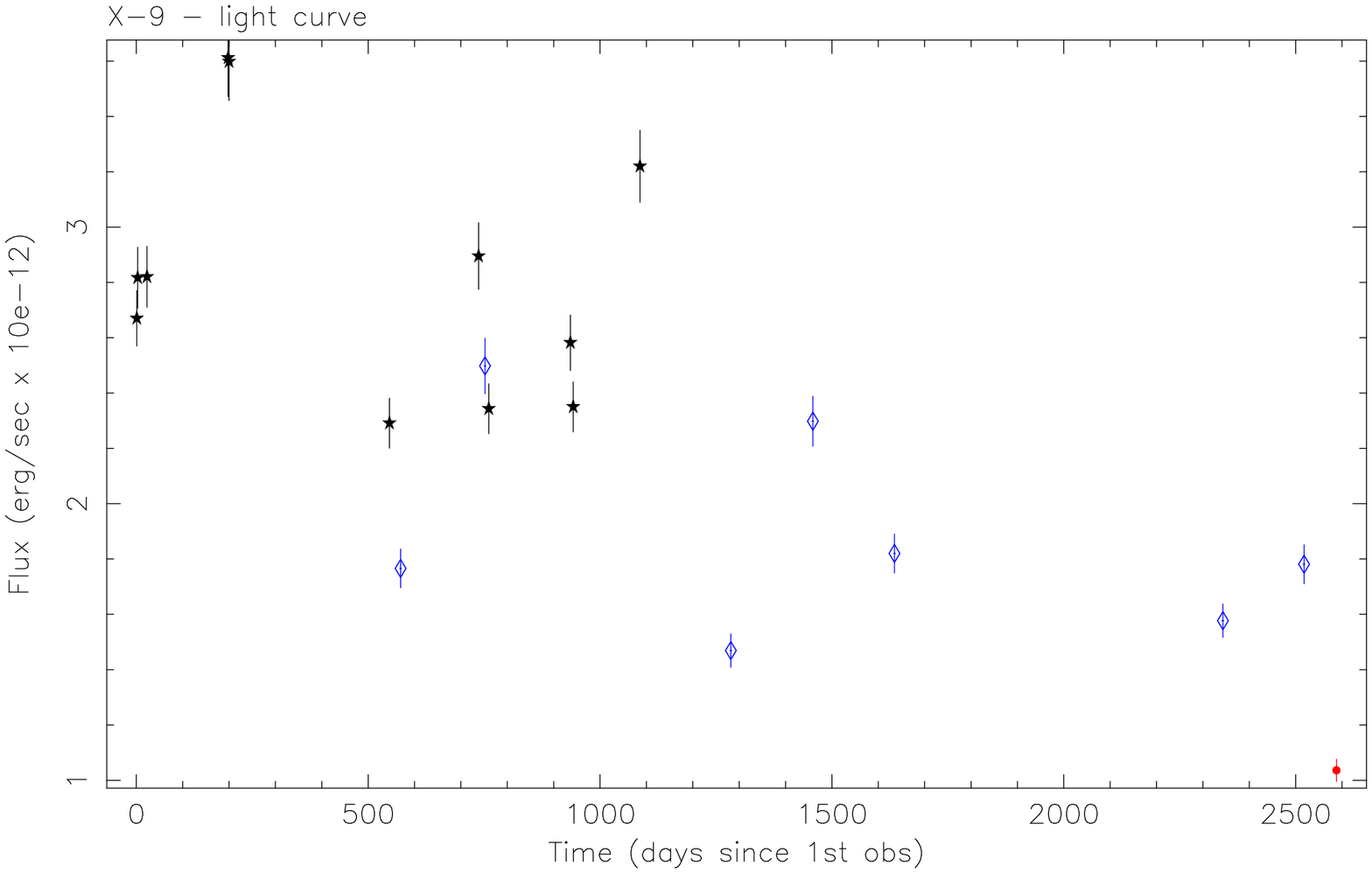,width=9.3cm,angle=270}}
\caption{Light curve of X-9. Each point is the average flux in one
observation. Same notation as in Fig. 1. Also this source shows variability on
several time scales.}
\label{x9flux}
\end{figure}

\noindent
{\bf X-9}\\
X-9 is a very bright X-ray source 12.5 arcmin away from M81 nucleus. Its
optical counterpart has not been clearly identified yet, and the nature of its
emission is still unclear. X-9 is strongly variable, as evident from the
light curve in Fig 2. Its rapid variability suggests that the bulk of
the emission comes from a point-like source, even if from HRI images we found 
some hint at a diffuse emission. The spectrum appears to be very complex and
variable. We divided all data into two groups, or spectral phases, according 
to the hardness ratio values of each observation, analysing them separately. We
found that the best fit for both groups is a multi-temperature black body disk
(successfully used to model galactic black hole candidate- Makishima et al.,
2000). The temperature we find for the two data sets are only marginally
comatible (kT$_1=0.58\pm0.07$ keV for the soft phase, kT$_2=0.68\pm0.04$ keV 
for the hard phase), so confirming the spectral variability. We also see a soft
excess in the spectrum, a possible signature of an extended component.\\

This work was supported in
part by NASA grant NAG5-2946 and NASA contract NAS8-39073(CXC) and in part by
MURST. This research has made use of the HEASARC online database and of the ESO
online DSS.

\end{document}